\pdfoutput=1

\documentclass[11pt, a4paper]{article}
\usepackage{geometry}
\usepackage{comment}
\usepackage{amsmath}
\usepackage{amsfonts}
\usepackage{amssymb}
\usepackage{graphicx}
\usepackage{mathrsfs}
\usepackage{latexsym}
\usepackage{color}
\usepackage{slashed, cancel}
\usepackage{hyperref}
\usepackage{bbold}
\usepackage{soul} 
\usepackage{booktabs}
\usepackage[font=footnotesize,labelfont=bf]{caption} 
\usepackage{tikz}
\usepackage{subcaption}

\usetikzlibrary{positioning,arrows,patterns}
\usetikzlibrary{decorations.pathmorphing}
\usetikzlibrary{decorations.markings}
\usetikzlibrary{calc}
\usetikzlibrary{shapes.misc}
\tikzset{
    photon/.style={decorate, decoration={snake}, draw=black, thick},
    fermionnoarrow/.style={draw=black, postaction={decorate}, thick},
    scalar/.style={draw=black, postaction={decorate}, decoration={markings,mark=at position .55 with {\arrow{>}}}, thick, dashed},
    scalarnoarrow/.style={draw=black, postaction={decorate},  thick, dashed},
    fermion/.style={draw=black, postaction={decorate},decoration={markings,mark=at position .55 with {\arrow{>}}}, thick},
    gluon/.style={decorate, draw=black, decoration={coil,amplitude=4pt, segment length=5pt}, thick},
    vertex/.style={draw,shape=circle,fill=black,minimum size=3pt,inner sep=0pt},
    effvertex/.style={draw,shape=circle,pattern=north west lines,pattern
    color=black,minimum size=6pt,inner sep=0pt},
    cross/.style={cross out, draw=black,thick, minimum size=6pt, inner sep=0pt, outer sep=0pt}
}

\usepackage{enumitem}

\usepackage{journals}

\usepackage[square, numbers, comma, sort&compress]{natbib} 
\hypersetup{urlcolor=blue, colorlinks=true} 

\usepackage{fancyhdr}
\newcommand{\iu}{{i\mkern1mu}}

\numberwithin{equation}{section}

\definecolor{rossos}{rgb}{0.8,0.2,0.3}
\definecolor{bluscuro}{rgb}{0.15, 0.2, .85}
\definecolor{bluchiaro}{cmyk}{1,.3,0.,0.1}
\hypersetup{colorlinks, citecolor=bluscuro, linkcolor=bluscuro, urlcolor=bluscuro}

\makeatother   
\newcommand{\GeV}{{\rm \,GeV}}
\newcommand{\TeV}{{\rm TeV}}

\newcommand{\ket}[1]{\ensuremath{\left|#1\right\rangle}}
\newcommand{\braket}[1]{\ensuremath{\left\langle#1\right\rangle}}

\def\tr{\textrm{Tr}}

 \def\be   {\begin{equation}}   \def\ee   {\end{equation}}
 \def\ba   {\begin{array}}      \def\ea   {\end{array}}
 \def\bea  {\begin{eqnarray}}   \def\eea  {\end{eqnarray}}
 \def\bean {\begin{eqnarray*}}  \def\eean {\end{eqnarray*}}

\begin{document}

\vspace{0.5cm}
\begin{center}

{\LARGE \textbf {
		Unitarisation of EFT Amplitudes for Dark Matter Searches at the
		LHC
}}
\\ [1.5cm]

{\large
\textsc{Nicole F.~Bell}$^{\rm a,}$\footnote{\texttt{n.bell@unimelb.edu.au}},
\textsc{Giorgio Busoni}$^{\rm a,}$\footnote{\texttt{giorgio.busoni@unimelb.edu.au}},
\textsc{Archil Kobakhidze}$^{\rm b,}$\footnote{\texttt{archil.kobakhidze@sydney.edu.au}},
\textsc{David M.~Long}$^{\rm c,}$\footnote{\texttt{dlon0938@uni.sydney.edu.au}},
\textsc{Michael A.~Schmidt}$^{\rm b,}$\footnote{\texttt{michael.schmidt@sydney.edu.au}},
}
\\[1cm]

\large{
$^{\rm a}$ 
\textit{ARC Centre of Excellence for Particle Physics at the Terascale, School of Physics,
The University of Melbourne, Victoria 3010, Australia}\\
\vspace{1.5mm}
$^{\rm b}$ 
\textit{ARC Centre of Excellence for Particle Physics at the Terascale, School of Physics,
The University of Sydney, NSW 2006, Australia}\\
\vspace{1.5mm}
$^{\rm c}$ 
\textit{School of Physics,
The University of Sydney, NSW 2006, Australia}
}
\end{center}

\vspace{0.5cm}

\begin{center}
\textbf{Abstract}
\begin{quote}
We propose a new approach to the LHC dark matter search analysis
within the effective field theory framework by utilising the
$K$-matrix unitarisation formalism. This approach provides a
reasonable estimate of the dark matter production cross section at
high energies, and hence allows reliable bounds to be placed on the
cut-off scale of relevant operators without running into the problem
of perturbative unitarity violation. We exemplify this procedure for the
effective operator D5 in monojet dark matter searches in the
collinear approximation. We compare our bounds to those obtained using
the truncation method and identify a parameter region where the
unitarisation prescription leads to more stringent bounds.
\end{quote}
\end{center}

\def\thefootnote{\arabic{footnote}}
\setcounter{footnote}{0}
\pagestyle{empty}

\newpage
\pagestyle{plain}
\setcounter{page}{1}

\section{Introduction}

A dedicated search for Dark Matter (DM) at the Large Hadron Collider is
currently one of the foremost objectives in particle physics. 
The most generic search channel is the mono-jet plus missing
transverse energy signal, which searches for a single jet recoiling
against the momentum of the DM particles which escape the detector
unseen \citep{ATLAS:2012ky,Chatrchyan:2011nd,Aad:2015zva,Khachatryan:2014rra,Aaboud:2016tnv,CMS:2015jdt}. 
In order to make such a search possible, it is necessarily to have a
framework in which to describe the interactions of dark matter
particles with SM fields.  Given the plethora of possible dark matter
models in the literature, it is impractical to perform a dedicated
analysis of each model.  It is thus imperative to work with a small
number of models that capture the essential aspects of the physics in
some approximate way.  Effective field theories (EFTs) achieve this
aim, by parameterising the DM interactions with SM particles by a small
set of non-renormalizable operators.  For instance, the lowest order
operators that describe the interaction of a pair of fermionic DM
particles, $\chi$, with a pair of SM fermions, $f$, are of the form
\begin{equation}
\frac{1}{\Lambda^2}\left(\overline\chi \Gamma_\chi \chi\right) \left(\overline{f} \Gamma_f f \right),
\end{equation}
where the Lorentz structure $\Gamma_{\chi,f}$ can be
$1, \gamma_5, \gamma_\mu, \gamma_5\gamma_\mu, \sigma_{\mu\nu}$.  A
full set of operators can be found
in \citep{Goodman:2010ku,Goodman:2010yf}, where a standard naming
convention has been defined.  Such operators are not intended to be
complete description of DM interactions, valid at arbitrarily high
energy.  They would be obtained as a low energy approximation of some
more complete theory by integrating out heavy degrees of freedom.  The
energy scale $\Lambda$ is related to the parameters of that high
energy theory as $\Lambda = g/M$, where $g$ is a coupling constant and
$M$ is the mass of a heavy mediator.

The EFT description will clearly break down at energies comparable to
$\Lambda$, at which scale we expect the mediators to be produced
on-shell or give rise to cross section resonances.  Moreover, while
the EFT will provide physically well-behaved cross sections at low
energies, they will give rise to bad high energy behaviour if used
outside their region of validity. This manifests as a violation of
perturbative unitarity~\cite{Endo:2014mja, Shoemaker:2011vi,
Hedri:2014mua, Yamamoto:2014pfa}.  While these issues may be remedied
with a Simplified Model~\cite{Abercrombie:2015wmb} in which a mediator
is explicitly introduced, issues of unitarity violation can persist if
gauge invariance is not respected.  The shortcoming of EFTs and
Simplified Models that violate SM gauge
invariance~\cite{Bell:2015sza,Bell:2015rdw,Haisch:2016usn,Englert:2016joy}
or dark-sector gauge invariance~\cite{Kahlhoefer:2015bea,Bell:2016fqf}
have recently been discussed.

Given the usefulness of the EFT and Simplified Model description of DM 
interactions, they will continue to be used in collider DM search 
analyses. Therefore, it is important to limit analyses to parameters 
that respect perturbative unitarity.  One such approach is to use a 
truncation 
technique~\citep{Busoni:2013lha,Busoni:2014sya,Busoni:2014haa}, which 
introduces a momentum cutoff equal to the mass of the would-be 
integrated-out mediator.  In this paper we will instead use a procedure 
known as $K$-matrix unitarisation ~\cite{Wigner:1946zz,Wigner:1947zz,Chung:1995dx,Chung:2015sf,Jacob:1959at,Martin:1970ept} to enforce unitarisation of 
all scattering amplitudes.  Although this procedure will not capture the 
resonance structure of the true high energy theory, it will force 
scattering amplitudes to be well behaved at high energies, allowing us 
to derive meaningful limits on EFT models from LHC collisions with 
high centre of mass energies.

We will use the $K$-matrix approach to unitarise the 2 to 2 scattering
amplitudes, such as $\overline{q}q\rightarrow \overline\chi\chi$.
This will allow us to determine unitarised cross sections for the 2 to
3 mono-jet processes such as
$\overline{q}q\rightarrow \overline\chi\chi g$, under the assumption
that the gluon can be treated with the collinear approximation.  We
will also compare the results obtained from this unitarisation
technique with those obtained with truncation.
The rest of the paper is organised as follows: in
Section~\ref{sec:unitar} we summarise the theoretical framework for
the unitarisation procedure. We illustrate the unitarisation procedure
in two toy models in Section \ref{sec:toymodels} and apply it to the
standard vector operator D5 in
Section \ref{sec:D5}. Section \ref{sec:conclusions} contains the
conclusions, while in Appendix \ref{app:collinear} we derive the
relevant cross sections in the collinear limit.


\section{K-Matrix Unitarisation}\label{sec:unitar}

The $K$-matrix formalism was first introduced in
Ref.~\cite{Wigner:1946zz,Wigner:1947zz}. It is a technique to impose
unitarity on amplitudes which naively violate unitarity. In the
derivation we largely follow the notation and arguments in
Refs.~\cite{Chung:1995dx,Chung:2015sf,Kilian:2014zja}~\footnote{See
Ref.~\cite{Jacob:1959at,Martin:1970ept} for further details.}.
Unitarity of the $S$-matrix,
\begin{equation}
	S=\mathbb{I} + 2 \iu T\;,
\end{equation}
implies the well-known relation for the $T$-matrix
\begin{align}
	T-T^\dagger  &= 2 \iu T T^\dagger\;.
\end{align}
Note the factor of $2$ in the definition of the $T$-matrix which has been introduced for convenience.

Following the seminal work by Jacob and Wick~\cite{Jacob:1959at}, for scattering processes $a\,b\to c\,d$ we can describe both the initial and the
final state in terms of two-particle helicity states $\ket{\Omega \lambda_1
\lambda_2}$ which are characterised by the helicities $\lambda_i$ of the two
particles and two angles $\theta$ and $\phi$, collectively denoted $\Omega$.
Choosing the initial state to align with the $z$-axis, the individual $T$-matrix
element for a process $a\,b \to c\,d$ with fixed helicities in the initial and
final state is given by
\begin{equation}
        \braket{\Omega\lambda_c\lambda_d|T|0\lambda_a\lambda_b} = \frac{1}{4\pi}
        \sum_J \left(2J+1\right) T^J_{\lambda^\prime\lambda}
        \mathcal{D}^{J*}_{\lambda\lambda^\prime}(\phi,\theta,0)\;,
\end{equation} 
in terms of the partial waves 
\begin{equation}
	T^J_{\lambda^\prime\lambda}\equiv\braket{J\lambda_c\lambda_d|T|J\lambda_a\lambda_b}
	= \int d\Omega \braket{\Omega\lambda_c\lambda_d|T|0\lambda_a\lambda_b}
	\mathcal{D}^J_{\lambda\lambda^\prime}(\phi,\theta,0)  \;,
\end{equation}
the Wigner $D$-functions
$\mathcal{D}^J_{\lambda\lambda^\prime}$ with total angular momentum $J$, and the
resultant helicity of the two-particle states $\lambda=\lambda_a-\lambda_b$ and
$\lambda^\prime=\lambda_c-\lambda_d$, where we used the normalisation of the Wigner $D$-functions in Ref.~\cite{Chung:2015sf}.
Assuming that no three-particle states are kinematically accessible, an analogous unitarity relation holds for each partial wave $T^J_{\lambda\lambda^\prime}$ separately,
\begin{equation}
	T^J - T^{J\dagger} = 2\iu T^{J\dagger}T^J,
\end{equation}
in terms of matrices $T^J$ with components $T^J_{\lambda^\prime\lambda}$.
This condition can be rewritten in terms of 
\begin{align}\label{eq:defK}
		\left(K^J\right)^{-1}\equiv \left(T^{J}\right)^{-1} + \iu\mathbb{I} =
	\left(\left(T^J\right)^{-1} + \iu\,
\mathbb{I}\right)^\dagger\;, 
\end{align}
which motivates the definition of the $K$-matrix for the J\textsuperscript{th} partial wave,
$K^J$. The $K$-matrix is hermitean, $K^J=K^{J\dagger}$. 
If the $S$-matrix is invariant under time reversal, the $K$-matrix is symmetric
and thus $K^J$ and $\left(K^J\right)^{-1}$ are real. Hence $(K^J)^{-1}$ can be
considered as the real part of $(T^J)^{-1}$ and the imaginary part of $T^J$ is
determined by the term $\iu\mathbb{I}$ in Eq.~\eqref{eq:defK}. We can invert the
relation in Eq.~\eqref{eq:defK} to obtain
\begin{equation}
	T^J\equiv \frac{1}{(K^J)^{-1} -\iu\mathbb{I}}\;.
\end{equation}%
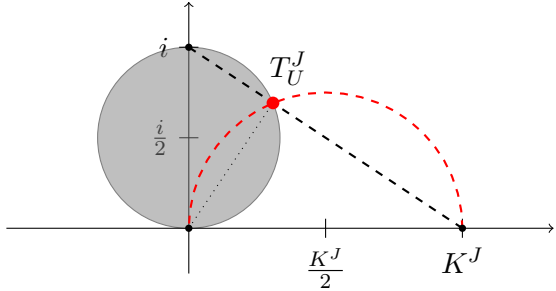
\begin{figure}[bt]\centering
	\begin{tikzpicture}[scale=1.2]
	\draw[->] (-2,0) -- (4,0);
	\draw[->] (0,-0.5) -- (0,2.5);
	\draw (1.5,3pt) -- (1.5,-3pt)   node [below]  {$\tfrac{K^J}{2}$};
	\draw (3,3pt) -- (3,-3pt)   node [below] {$K^J$};
	\draw (3pt,2) -- (-3pt,2)   node [left]   {$i$};
    \draw (3pt,1) -- (-3pt,1)   node [left] {$\tfrac{i}{2}$};
    \path [draw=black,fill=gray,semitransparent] (0,1) circle (1);
     \draw [dashed,red,thick,domain=0:180] plot ({1.5+1.5*cos(\x)}, {1.5*sin(\x)});
    \draw [dashed,thick] (3,0) -- (0,2);
    \draw [dotted] (0,0) -- ({12/13,18/13}) node [above,yshift=1mm,xshift=2mm] {$T_U^J$};
    \path [fill=red]  ({12/13,18/13}) circle (0.07);
    \path [fill=black]  (3,0) circle (0.04);
    \path [fill=black]  (0,2) circle (0.04);
    \path [fill=black]  (0,0) circle (0.04);
\end{tikzpicture}
\caption{Argand circle and Thales projection.}
\label{fig:argand}
\end{figure}%
The matrix $T^J$ is given by the stereographic projection of the $K$-matrix on
the Argand circle as shown in Fig.~\ref{fig:argand}. 
If perturbative unitarity is violated in any amplitude, it can be enforced by imposing reality on
$(K^J)^{-1}$, i.e. replacing $(K^J)^{-1}$ by Re$[(T^J)^{-1}]$, which leads to
the unitarised $T$-matrix\footnote{Note that this prescription is not analytic
at $T^J=0$ ~\cite{Kilian:2014zja}.
$K$-matrix unitarisation does not enforce a consistent analytic structure ~\cite{Truong:1991gv}. In practice this is not important, because we are interested in studying monojet searches at the LHC where the amplitudes are large and $T^J\neq 0$. }
\begin{align}
	T_{U}^J\equiv \frac{1}{\mathrm{Re}\left[(T^J)^{-1}\right] -\iu\mathbb{I}}\;.
\label{eq:unitarization}
\end{align}
Particularly in case the $T$-matrix quadratically grows with the centre of mass energy,
$T^J \propto \tfrac{s}{16\pi\Lambda^2}$, the unitarised $T$-matrix asymptotically reaches saturation
\begin{equation}
	T_U^J = \frac{1}{\tfrac{16\pi\Lambda^2}{s}- \iu}
	\stackrel{s\to\infty}{\longrightarrow} \iu\;,
\end{equation}
which can be interpreted as a resonance at infinity.
Note that the restriction to the real part of $\left(T^J\right)^{-1}$
can be understood as the Thales projection onto the real
axis~\cite{Kilian:2014zja}, if the $T$-matrix $T^J$ is complex, i.e. points lying on the red dashed circle in
Fig.~\ref{fig:argand} are projected onto the same unitarised $T$-matrix $T^J_U$
as $K^J$. All discussed operators in Secs.~\ref{sec:toymodels} and \ref{sec:D5}
lead to a real $T$-matrix $T^J$ in the considered scattering processes.
Alternatively, following Ref.~\cite{Gupta:1949rh,Gupta:1977pd} the hermitean $K$-matrix
can be considered as an approximation to the scattering amplitude, which can be
obtained order by order in perturbation theory using Eq.~\eqref{eq:defK}. Using
the fact that the $K$-matrix is the Cayley
transform of the $S$-matrix~\cite{Heitler:1941xx,Schwinger:1948yk}
\begin{equation}\label{eq:Cayley}
	S=\frac{\mathbb{I}+\iu K}{\mathbb{I}-\iu K}\;,
\end{equation}
it is possible to reconstruct a unitary $S$-matrix starting from an approximate
$K$-matrix. The $S$-matrix defined in Eq.~\eqref{eq:Cayley} restores unitarity,
which is lost in the usual expansion of the $S$-matrix, if only a finite number
of terms are taken into account in perturbation theory.
The $K$-matrix formalism can be considered minimal, since it does not introduce new parameters
or visible structures in scattering amplitudes like resonances. However it
does not yield a viable UV completion of the effective theory. New resonances
have to be included by hand. See Refs.~\cite{Alboteanu:2008my,Kilian:2014zja,Delgado:2015kxa} for a
recent discussion in the context of $WW$ scattering.

In the following, we will make use of this prescription to obtain unitary
amplitudes for DM pair production at the LHC.
Taking the normalisation of the two-particle states properly into account, the $T$-matrix is related to the usual Lorentz-invariant matrix element $\mathcal{M}_{fi}$ by 
\begin{equation}
	\braket{\Omega\lambda_c\lambda_d|T|0\lambda_a\lambda_b} =
\frac{1}{32\pi^2} \sqrt{\frac{4 p_f p_i}{s}} \mathcal{M}_{fi}\;,
\end{equation}
and analogously the partial waves.  In the ultra-relativistic limit,
the initial and final state phase space densities $2p_{i,f}/\sqrt{s}$
approach unity, simplifying the calculation of the unitarised
$T$-matrix considerably.  Finally, the differential cross section in
terms of the $T$-matrix element is given by
\begin{equation}
	\frac{d\sigma_{fi}}{d\Omega} = \frac{(4\pi)^2}{s} \frac{s}{4\,p_i^2} \left|
\braket{\Omega\lambda_c\lambda_d|T|0\lambda_a\lambda_b}\right|^2\;,
\end{equation}
and thus the total cross section can be conveniently expressed in terms of the
partial waves
\begin{equation}
	\sigma_{fi} =\frac{4\pi}{s} \frac{s}{4\,p_i^2}  \sum_J (2J +1)
	\left|T_{\lambda^\prime\lambda}^J\right|^2 = \frac{4\pi}{s-4m_i^2}
	\sum_J (2J+1) \left|T_{\lambda^\prime\lambda}^J\right|^2\;.
\end{equation}
Note that this is the cross section for fixed helicities. 
The unpolarised and color averaged cross section is obtained in the usual way by
averaging over the initial state helicities and number of colours and summing over the final state ones, i.e.,
\begin{equation}
	\sigma(q\bar q \to X) = \frac{1}{12} \sum_\mathrm{helicities} \sigma_{fi}
\end{equation}
for the unpolarised cross section $q\bar q \to X$ with two quarks in the initial
state. 
The unitarised cross section is obtained by replacing
$T_{\lambda^\prime\lambda}^J$ by the corresponding unitarised $T$-matrix element
$T^J_{U\lambda^\prime\lambda}$. Thus the cross section is unitarised for each
quark color and helicity separately.

\section{Simple Two-Channel Models}\label{sec:toymodels}
To illustrate the unitarisation procedure, we will make a simplifying
assumption concerning the quark states in the operator and consider
two simple models which feature only two channels.  The effective
operator D5 shall then be discussed in the next section.

\subsection{States}
As we are working in the collinear approximation, in the $T$-matrix we
ought to consider all coupled two-particles states to expect the
unitarity of the $S$-matrix to hold. If we consider the SM 
plus the 
DM particle coupled with an
EFT operator, this implies the consideration of all possible
two-particle states in the standard model with zero charge, baryon and
lepton number, in addition to $\chi
\bar{\chi}$. Taking into account color, helicity and flavour, this results in $3\cdot 3 \cdot 4 \cdot 6 = 216$ states for the quarks alone. To simplify the framework, we consider only the singlet color state
\begin{equation}
\frac{R\bar{R}+V\bar{V}+B\bar{B}}{\sqrt{3}}\;,
\end{equation}
because all other color combinations decouple from this state and the DM sector.
Moreover we assume the same operator suppression scale $\Lambda$ for all quark flavours. In this case we can also consider just one flavour state:
\be
\frac{u\bar{u}+d\bar{d}+s\bar{s}+c\bar{c}+b\bar{b}+t\bar{t}}{\sqrt{6}}\;,
\ee
as, again, all other flavour combinations decouple from this state and the DM sector.
Now, if we ``turn off" electro-weak interactions, i.e. approximating $\alpha_{EW}\ll \alpha_s$, this state decouples from all other standard model states, and only couples to itself and the DM states.

\subsection{EFT Motivated by T-channel Scalar Exchange}
\label{sec:toy1}
We now consider a toy model scenario that can be solved analytically.
We take the following
EFT operator connecting the dark and the visible sector:
\begin{equation}
	\mathcal{L}_1=\frac{1}{\Lambda_{q\chi}^2}\bar{q}\gamma_\mu P_R
	q\bar{\chi}\gamma^\mu P_L\chi\;.
\label{eq:qchiop}
\end{equation}
This operator can arise by integrating out a heavy coloured scalar
t-channel mediator
coupling only to right-handed quarks and left-handed DM particles. 
In the limit of massless particles, $s\gg m_\chi^2, m_q^2$, the only non-zero
$T$-matrix elements are $\braket{\chi_L\bar\chi_R|T|q_R \bar
	q_L}$\footnote{Note that right-handed (left-handed) particles have helicity
$+\frac12$ ($-\frac12$).} and the matrix element
$\braket{q_R\bar q_L|T|\chi_L\bar \chi_R}$ related by time-reversal. 
Thus we are left with a $2 \times 2$ $T$-matrix,
\begin{equation}
	T= - \frac{1}{16\pi^2} \frac{s}{\Lambda_{q\chi}^2}\; 
	\begin{pmatrix}
		0 & 1 \\ 
	1 & 0 \\
\end{pmatrix}
\sin^2\frac\theta2\;, 
\end{equation}
in the basis of the two helicity $1$ two-particle states $\left(\ket{q_R\bar q_L}, \ket{\chi_L\bar\chi_R}\right)$.
We only include the contribution of the effective operator and neglect any QCD
contribution. The partial wave expansion only contains the term with total
angular momentum $J=1$, 
\begin{align}
	T^1&=
		-\frac{1}{12\pi} \frac{s}{\Lambda_{q\chi}^2} \begin{pmatrix}
0   &   1 \\
1 &  0  \\
\end{pmatrix}\;,
\end{align}
which grows linearly with $s$ and thus is going to violate
perturbative unitarity for scales $s\gtrsim 12\pi\Lambda_{q\chi^2}$. 
After unitarising the amplitudes using $K$-matrix unitarisation, the unitarised
amplitude turns out to be 
\begin{align}\label{eq:T1unit}
	T_U^1&=\frac{1}{s^2+144\pi^2\Lambda_{q\chi}^4} \begin{pmatrix}
	\iu s^2   &  - 12\pi s \Lambda_{q\chi}^2   \\
	- 12\pi s \Lambda_{q\chi}^2 &  \iu s^2  \\
\end{pmatrix}\;.
\end{align}
Note that the unitarisation procedure introduces contributions to the
scattering of $\bar q q \to \bar q q$ and
$\bar \chi\chi \to \bar \chi\chi$. The denominator leads to a smooth
cutoff around $s\sim 12\pi \Lambda_{q\chi}^2$, indicating that the
non-unitarised amplitude strongly violates perturbative unitarity
above such energy. When discussing the validity of the EFT, this in
turn means that, unless new states and/or new interactions are
introduced, the EFT breaks at this energy scale.  The unitarised
$T$-matrix is well-behaved for large $s$ and converges to
$\iu\mathbb{I}$ and it can be thus used to interpret scattering
events, like monojet signatures at the LHC. In fact, the high-energy
tail leads to a negligible contribution due to the suppression of the
parton distribution function at high-energy in contrast to the EFT.

\subsection{EFT Motivated by S-Channel Vector Boson Exchange}
Generally there might also be operators between two quark currents or two dark
matter currents. As second example we consider an effective theory with 
three operators
\begin{equation}
	\mathcal{L}_{2} = \frac{1}{2\Lambda_{qq}^2}\bar{q}\gamma_\mu P_R
	q\bar{q}\gamma^\mu P_R q+ \frac{1}{\Lambda_{q\chi}^2}\bar{q}\gamma_\mu
	P_R q\bar{\chi}\gamma^\mu P_R\chi +
	\frac{1}{2\Lambda_{\chi\chi}^2}\bar{\chi}\gamma_\mu P_R
	\chi\bar{\chi}\gamma^\mu P_R\chi\;,
\label{eq:allop}
\end{equation}
which might arise from a Simplified Model with a $Z^\prime$ gauge
boson coupling only to the right-handed quark and DM currents. 
In such a model the EFT parameters are related to those of the UV complete theory according to $\Lambda_{qq}^2=M_{Z'}^2/g_q^2$, $\Lambda_{q\chi}^2=M_{Z'}^2/(g_qg_\chi)$ and $\Lambda_{\chi\chi}^2=M_{Z'}^2/g_\chi^2$, where $g_{q,\chi}$ are the couplings of the $Z'$ to the quarks and the DM, and $M_{Z'}$ is the mediator mass.
The effective operators lead to four non-vanishing entries in the
$T$-matrix, $\braket{q_R\bar q_L|T|q_R\bar q_L}$,
$\braket{\chi_R\bar\chi_L|T|\chi_R\bar\chi_L}$,
$\braket{\chi_R\bar\chi_L|T|q_R\bar q_L}$, and $\braket{q_R\bar
q_L|T|\chi_R\bar \chi_L}$, where the latter two are related by
time-reversal. The $T$-matrix in the basis $\left( \ket{q_R \bar
q_L}, \ket{\chi_R\bar \chi_L}\right)$ is then given by
\begin{equation}
	T=-\frac{1}{16\pi^2} 
	\begin{pmatrix}
		 \frac{2s}{\Lambda_{qq}^2}
		  &\frac{s}{\Lambda_{q\chi}^2}
		\\
		\frac{s}{\Lambda_{q\chi}^2}
&\frac{2s}{\Lambda_{\chi\chi}^2}
	\end{pmatrix} \cos^2\frac\theta2
	\;.
\end{equation}
Gluon s-channel exchange between quark - anti-quark pairs leads to an additional
contribution to the $\braket{q_R\bar q_L|T|q_R\bar q_L}$ element. It does not
grow with $s$ like the other contributions and thus can be neglected for large
$s$, when perturbative unitarity becomes an issue. The only non-vanishing term
in the partial wave expansion has total angular momentum $J=1$ reading
\begin{align}
	T^1& =-\frac{1}{12 \pi}
		\begin{pmatrix}
		 \frac{2s}{\Lambda_{qq}^2}
		  &\frac{s}{\Lambda_{q\chi}^2}
		\\
		\frac{s}{\Lambda_{q\chi}^2}
&\frac{2s}{\Lambda_{\chi\chi}^2}
	\end{pmatrix}\;.
\end{align}
The expression for the unitarised $T$-matrix turns out to be complicated. 
Assuming an underlying Simplified Model with a $Z^\prime$ mediator, the operator
suppression scales are related via
\begin{equation}
\Lambda_{qq}\Lambda_{\chi\chi}=\Lambda_{q\chi}^2\;.
\label{eq:constrain}
\end{equation}
This motivates the definition of the ratio
\begin{equation}\label{eq:ratio}
r=\frac{\Lambda_{q\chi}}{\Lambda_{\chi\chi}}=\frac{\Lambda_{qq}}{\Lambda_{q\chi}}\;.
\end{equation}
In terms of the ratio $r$, the unitarised $T$-matrix, $T^1$, is 
\begin{equation}
	T_{U,r}^1 =
	\frac{1}{r^2s^2 -8 \iu \pi  \left(r^4+1\right) s \Lambda _{q\chi}^2-48
	\pi ^2 r^2 \Lambda _{q\chi }^4}
	\begin{pmatrix}
	 \iu s^2 r^2 + 8 \pi s \Lambda _{q\chi}^2 & 4 \pi  r^2 s \Lambda _{q\chi}^2 \\
		4 \pi  r^2 s \Lambda _{q\chi}^2 & \iu s^2 r^2 + 8 \pi s \Lambda _{q\chi}^2 \\
	\end{pmatrix}\;.
\end{equation}
Note that one can always parameterize new physics using a complete set of EFT
operators like the ones in Eq. \eqref{eq:allop}, thus this choice is not model
dependent, if one chooses a complete basis. The only model-dependent hypothesis
we are using comes from imposing the relation \eqref{eq:constrain} based on the
assumption that the chosen EFT operators comes from an integrated-out
$Z^\prime$ mediator. Even though this choice is model dependent, we will keep this
constraint to reduce the number of parameters of the model. In the following we
will restrict ourselves to this relation for simplicity and study the impact of
the unitarisation procedure on the cross section using the well-studied D5
operator and the corresponding four-fermion operators with only quark and dark
matter fields, respectively.

\section{Unitarising the Effective Operator D5}\label{sec:D5}
The $K$-matrix unitarisation procedure can be applied to any of the
studied operators. We will focus on the operator D5\footnote{The
operator D5 belongs to the list of operators presented in
Ref.~\cite{Goodman:2010ku}, which have been widely used in the LHC
monojet searches reported by the ATLAS and CMS collaborations. See
Tab.~\ref{tab:opList} for the full list of operators.}  which might arise
from a Simplified Model with a $Z^\prime$ gauge boson coupling to both
the quark and DM vector currents. Besides the operator D5, whose
Wilson coefficient we denote by $\Lambda_{q\chi}^{-2}$, we have to
consider the two four-fermion operators with only quarks and DM
particles $\chi$, respectively
\begin{equation}\label{eq:D5}
	\mathcal{L}_{D5} = \frac{1}{2\Lambda_{qq}^2}\bar{q}\gamma_\mu 
	q\bar{q}\gamma^\mu q + \frac{1}{\Lambda_{q\chi}^2}\bar{q}\gamma_\mu
	 q\bar{\chi}\gamma^\mu \chi +
	\frac{1}{2\Lambda_{\chi\chi}^2}\bar{\chi}\gamma_\mu 
	\chi\bar{\chi}\gamma^\mu \chi\;.
\end{equation}
The explicit expressions for the $T$-matrix, the partial waves and the
unitarised partial waves are summarised in App.~\ref{app:D5}.
Similarly to the second toy model in the previous section, we assume relation
\eqref{eq:constrain} for simplicity and express the results in terms of the
ratio \eqref{eq:ratio}.
$K$-matrix unitarisation does not depend on this assumption, but it considerably simplifies the
analysis by constraining the parameter space of the three
Wilson coefficients to the two-dimensional submanifold defined by Eq.~\eqref{eq:constrain}.

Before comparing the result of $K$-matrix unitarisation with the 8 TeV ATLAS EFT limits
for the operator D5~\cite{Aad:2015zva} and the method of truncation, we comment
on the validity of the collinear approximation and the importance of quark jets.

\subsection{Validity of Collinear Approximation}
The collinear approximation is technically only valid in the limit of small
scattering angles, i.e. small transverse momentum $p_T$ compared to the centre
of mass energy $\sqrt{s}$. Thus it is essential to
estimate how well the collinear approximation performs for monojet searches,
which usually employ a high cut on $p_T$ to suppress QCD background. The full
three-body final state cross section for the effective operator D5 with an
emission of one gluon jet is presented in the appendix of
Ref.~\cite{Busoni:2014sya}.%
  \begin{figure}[!hbt]
    \centering
    \includegraphics[width=0.7\textwidth]{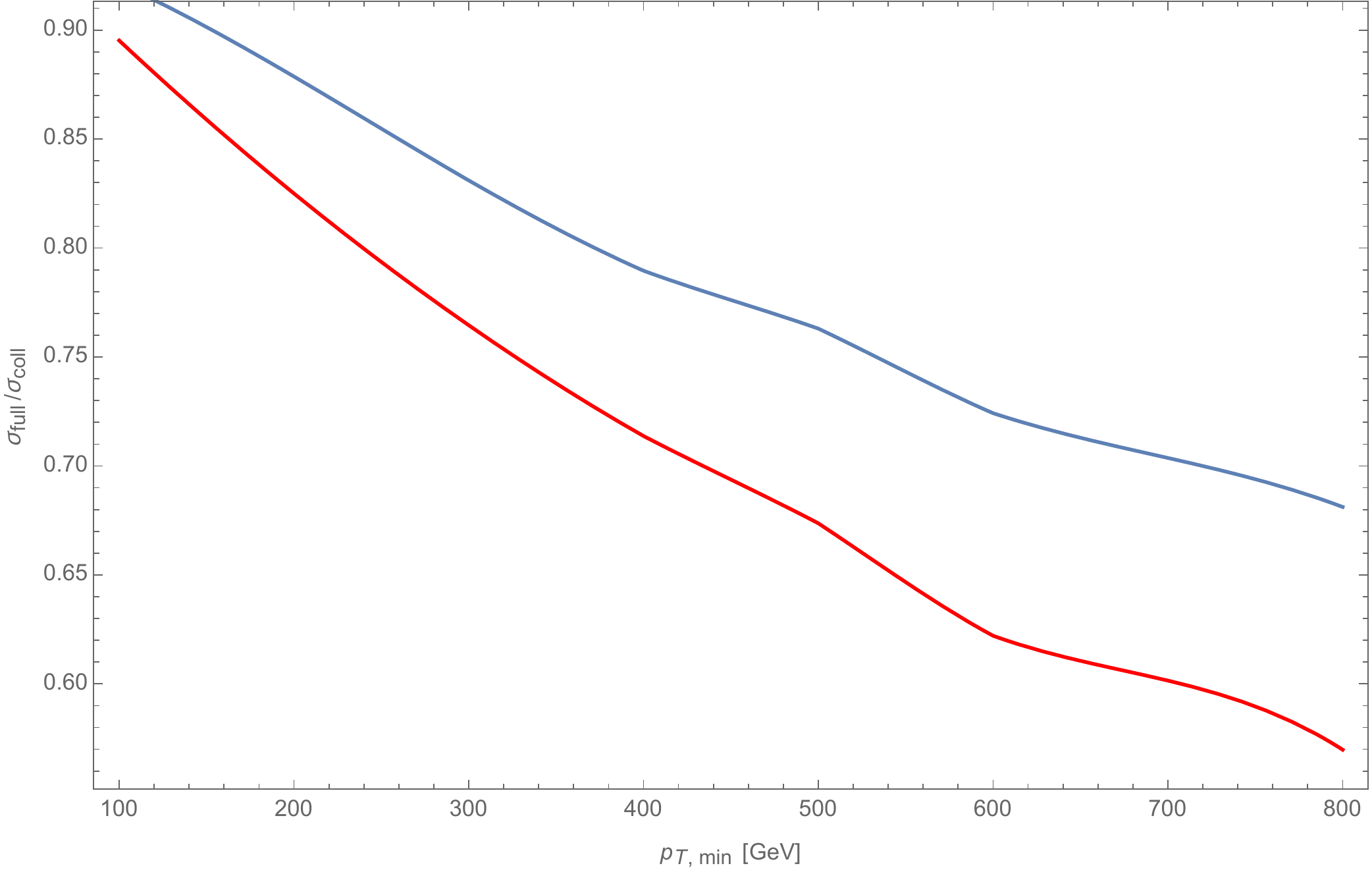}
    \caption{Ratio of the full cross section to the collinear one as a function
    of the minimum transverse momentum $p_T$ for $m_{DM}=100 \GeV$. The Blue
    line refers to beam energy of $13 \TeV$, the red one to $8 \TeV$.}
    \label{fig:collinear}
\end{figure}
Fig.~\ref{fig:collinear} depicts the ratio of the cross section using
the analytic result in Ref.~\cite{Busoni:2014sya} over the cross
section obtained in the collinear approximation as a function of the
minimum $p_{T,min}$ both for 8 TeV (red line) and 13 TeV (blue
line).\footnote{Note that the collinear limit for the effective
operators D1 and D4 agrees with the exact result.} The collinear
approximation leads to an enhancement of less than about 10\% of the
cross section for a minimum $p_{T,min}\simeq 100$ GeV, which grows to
$45\%$ ($30\%$) with $p_{T,min}=800$ GeV for 8 TeV (13 TeV) centre of
mass energy. The ATLAS 8 TeV monojet analysis~\cite{Aad:2015zva}
required $p_T > 120$ GeV and thus the collinear approximation
overestimates the cross section by about $13\%$. The 13 TeV monojet
searches plan to require $p_{T,min} =600$ GeV leading to about 37\%
overestimation of the cross section by taking the collinear limit.  We
expect similar results for the cross section in the unitarised
EFT, which is suggested by the fact that the cross
section in the effective theory can be factorised in the two-body
cross section $q\bar q\to
\chi\bar\chi$ and a function dependent on the scattering angle of the jet and
its rapidity. Consequently we expect the overestimation by taking the
collinear limit to mostly cancel out in the ratio of the cross
sections ($R_U$ and $R_\Lambda$, defined below).  Hence the
ratios calculated with the collinear approximation will be closer to
the values obtained from a full 3-body final state calculation than
the result in Fig.~\ref{fig:collinear} suggests. Thus the collinear
approximation works well, which is also supported by a similar
analysis in Ref.~\cite{Birkedal:2004xn}.  Going beyond the collinear
limit requires the inclusion of three-body states in the $T$-matrix
rendering the $K$-matrix unitarisation procedure more complicated. We
will defer an analysis beyond the collinear limit to a future
publication.

\subsection{Importance of Quark-Jets}
\begin{figure}[bth!]
	\centering
	\begin{subfigure}{0.45\textwidth}
		\begin{tikzpicture}[node distance=0.6cm and 1cm]
		\coordinate[label=left:$q$] (l1);
        \coordinate[effvertex, below right=of l1] (vEff);
	\coordinate[label=left:$q$,below left=of vEff] (l2);
	\coordinate[vertex] (vJet) at ($(l1)!0.5!(vEff)$);
	\coordinate[label=right:$g$, above right=of vJet] (j);
        \coordinate[label=right:$\chi$, above right=of vEff] (r1);    
        \coordinate[label=right:$\chi$, below right=of vEff] (r2);    
        \draw[fermion] (l1) -- (vJet) ;
	\draw[fermion] (vJet) -- (vEff) node[midway,swap,below left,yshift=1mm] {$q$};
        \draw[fermion] (vEff) -- (l2) ;
	\draw[fermion] (r1) -- (vEff);
        \draw[fermion] (vEff) -- (r2) ;
	\draw[gluon] (vJet) -- (j) ;
        \end{tikzpicture}
	\hspace{2ex}
\begin{tikzpicture}[node distance=0.6cm and 1cm]
        \coordinate[label=left:$q$] (l1);
        \coordinate[effvertex, below right=of l1] (vEff);
        \coordinate[label=left:$q$,below left=of vEff] (l2);
	\coordinate[vertex] (vJet) at ($(l2)!0.5!(vEff)$);
	\coordinate[label=right:$g$, below right=of vJet] (j);
        \coordinate[label=right:$\chi$, above right=of vEff] (r1);    
        \coordinate[label=right:$\chi$, below right=of vEff] (r2);    
        \draw[fermion] (l1) -- (vEff) ;
        \draw[fermion] (vEff) -- (vJet)  node[midway,swap,above
		left,yshift=-0.1cm] {$q$};
	\draw[fermion] (vJet) -- (l2); 
	\draw[fermion] (r1) -- (vEff);
        \draw[fermion] (vEff) -- (r2) ;
	\draw[gluon] (vJet) -- (j) ;
        \end{tikzpicture}
	\caption{Gluon jets}\label{fig:gluon-jet}
\end{subfigure}
	\hfill
	\begin{subfigure}{0.45\textwidth}
	\begin{tikzpicture}[node distance=0.6cm and 1cm]
        \coordinate[label=left:$g$] (l1);
        \coordinate[effvertex, below right=of l1] (vEff);
        \coordinate[label=left:$q$,below left=of vEff] (l2);
	\coordinate[vertex] (vJet) at ($(l1)!0.5!(vEff)$);
	\coordinate[label=right:$q$, above right=of vJet] (j);
        \coordinate[label=right:$\chi$, above right=of vEff] (r1);    
        \coordinate[label=right:$\chi$, below right=of vEff] (r2);    
        \draw[gluon] (l1) -- (vJet) ;
	\draw[fermion] (vJet) -- (vEff) node[midway,swap, below left,yshift=1mm] {$q$};
        \draw[fermion] (vEff) -- (l2) ;
	\draw[fermion] (r1) -- (vEff);
        \draw[fermion] (vEff) -- (r2) ;
	\draw[fermion] (j) -- (vJet)  ;
        \end{tikzpicture}
	\hspace{2ex}
\begin{tikzpicture}[node distance=0.6cm and 1cm]
        \coordinate[label=left:$q$] (l1);
        \coordinate[effvertex, below right=of l1] (vEff);
        \coordinate[label=left:$g$,below left=of vEff] (l2);
	\coordinate[vertex] (vJet) at ($(l2)!0.5!(vEff)$);
	\coordinate[label=right:$q$, below right=of vJet] (j);
        \coordinate[label=right:$\chi$, above right=of vEff] (r1);    
        \coordinate[label=right:$\chi$, below right=of vEff] (r2);    
        \draw[fermion] (l1) -- (vEff) ;
        \draw[fermion] (vEff) -- (vJet)  node[midway,swap,above
		left,yshift=-0.1cm] {$q$};
	\draw[gluon] (vJet) -- (l2); 
	\draw[fermion] (r1) -- (vEff);
        \draw[fermion] (vEff) -- (r2) ;
	\draw[fermion] (vJet) -- (j) ;
        \end{tikzpicture}
	\caption{Quark jets}\label{fig:quark-jet}
\end{subfigure}
		\caption{Initial state radiation leading to monojet signature
		in DM pair production at the LHC.}\label{fig:monojet}
\end{figure}
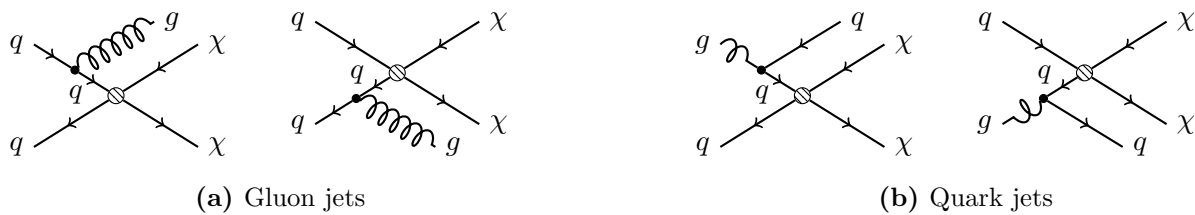

\begin{figure}[b!]
    \centering
    \includegraphics[width=0.7\textwidth]{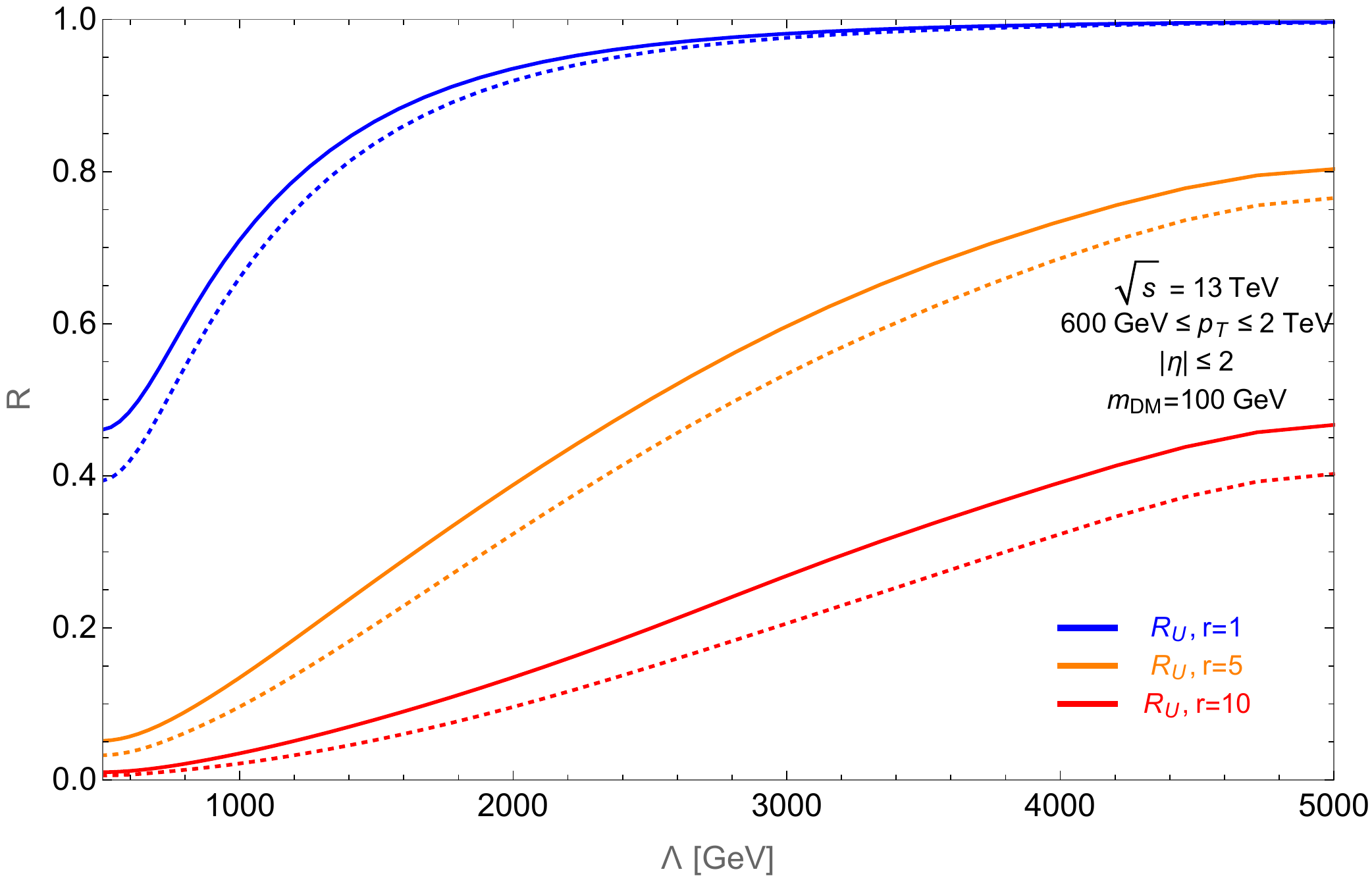}
    \caption{The ratio $R_{U}$ as a function of the cut-off scale $\Lambda$, for
    different values of $r$ for $m_{DM}=100 \GeV$. The solid lines refer to
    $R_{U}$ including both quark and gluon jets, the dotted lines refer to
    $R_{U}$ including only gluon jets.}
    \label{fig:B}
\end{figure}

In the previous subsection we only considered gluon jets, shown in
Fig.~\ref{fig:gluon-jet}, and neglected the
additional contribution from quark jets. It originates from diagrams with
gluons in the initial state as shown in Fig.~\ref{fig:quark-jet}.
Quark jets generally lead to a 10\% increase in the cross section, as it is
suggested by Fig.~6 in Ref.~\cite{Busoni:2014sya}. We included quark jets and
show in Fig.~\ref{fig:B} the ratio of the unitarised cross section over the cross
section using the effective field theory in the collinear limit for a fixed
value of the DM mass, $m_{DM}=100$ GeV,
\begin{align}
	R_U =
	\frac{\sigma_\mathrm{unitarised,coll.}}{\sigma_\mathrm{EFT,coll.}}\;,
\end{align}
for different values of $r=1,2,5$. The dotted lines show the ratio $R_U$, if quark jets
are neglected, while the solid lines take both contributions into account. The additional contribution of quark jets generally enhances the unitarised cross section over the EFT cross
section. 
 
\subsection{Reinterpretation of the 8 TeV ATLAS Monojet Limit}
ATLAS performed a monojet analysis with their full 8 TeV dataset of 20.3
fb$^{-1}$. The limits were interpreted for different EFT models including the
operator D5. Besides the EFT limit, ATLAS also quotes the limit obtained using
truncation, where only events are kept, which are consistent with the EFT
interpretation and satisfy the constraint
\begin{equation}
	\Lambda > \frac{Q_{tr}}{\sqrt{g_qg_\chi}} >
	2\frac{m_\mathrm{DM}}{\sqrt{g_qg_\chi}}\;,
\end{equation}
i.e. the requirement that the momentum transfer $Q_{tr}$ is always smaller than the mass
of the mediator $M = \sqrt{g_q g_\chi} \Lambda$, which is expressed in terms of
the cutoff scale $\Lambda$ and the couplings $g_{q,\chi}$ of the quarks and DM
particles $\chi$ to the mediator. In
case of D5, this could be the mass of an $Z^\prime$ gauge boson, which is
exchanged in the s-channel, and the corresponding gauge couplings with quarks
and DM. For gauge couplings, we naively expect the couplings to be of a
similar order of magnitude.
  \begin{figure}[!hbt]
    \centering
    \includegraphics[width=0.7\textwidth]{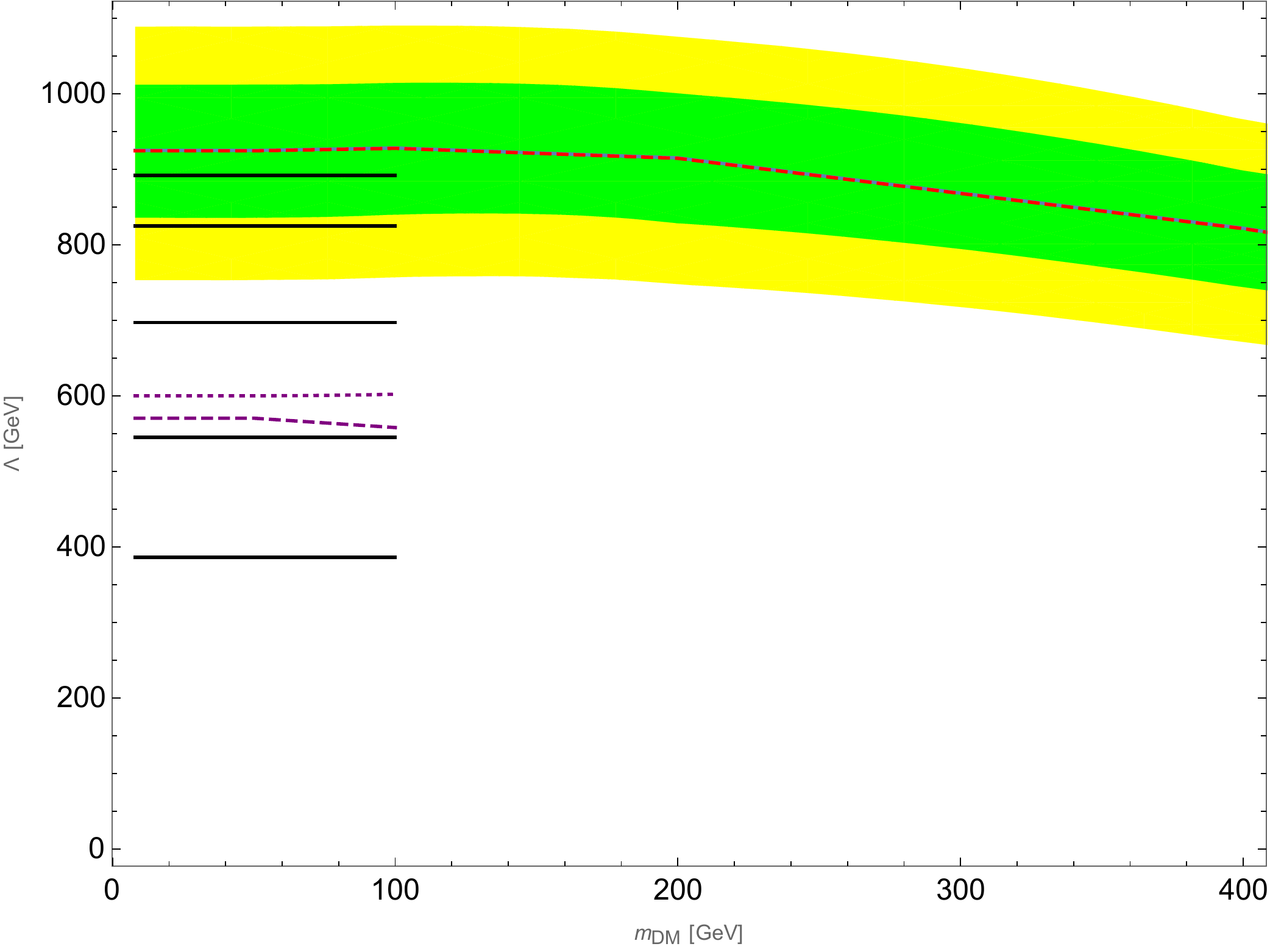}
    \caption{Reinterpretation of ATLAS limit at $8 \TeV$. The blue line refers to
    the ATLAS limit, the green and yellow band indicating the 1 and 2 sigma
    uncertainty bands, as in \citep{Aad:2015zva}. The red dashed line indicates
    the limit using truncation with maximal couplings, the purple dashed one
    using truncation with unit couplings. The purple dotted lines refers to our
    result using the collinear limit for the truncation with unit couplings, while the black lines refer to the unitarised amplitude with $r=1,2,3,4,5$ from top to bottom.}
    \label{fig:atlas8}
\end{figure}
We reproduce in Fig.~\ref{fig:atlas8} the official ATLAS 8 TeV monojet limit
shown in Fig.~10b of Ref.~\cite{Aad:2015zva}. The blue solid line refers to the
ATLAS EFT limit, and the green and yellow regions indicate the 1 and $2\sigma$
uncertainty bands. The red dashed line corresponds to the limit using truncation
with maximal couplings $g_qg_\chi=4\pi$ and the purple dashed line to the one
using truncation with  couplings $g_qg_\chi=1$. The purple dotted line is our
result for truncation with unit couplings using the collinear limit. The black
solid lines show the limit obtained using the unitarised amplitude with $r=1,2,3,4,5$ from top to
bottom. The limits are only shown for small DM masses $m_{DM}<100$ GeV,
because they are derived neglecting the DM mass. In our analysis, we
employ the collinear limit and only include the leading jet unlike the ATLAS analysis, which included a second
jet. These effects go in the opposite direction and partly cancel each other.
The unitarised amplitude with $r\leq3$ leads to a stronger limit than using
truncation with $g_qg_\chi=1$. 

\subsection{Future Projection to 13 TeV and Comparison to Truncation}

Using the cross section ratio, it is straightforward to apply the same method to
a future analysis. The EFT cross section is suppressed by the fourth power of
the scale of the effective operator $\Lambda\equiv\Lambda_{q\chi}$. Thus a
reduction of the unitarised cross section by a factor $R_U$ approximately
results in a decrease of the limit on the scale $\Lambda$ by a factor of
$R_U^{1/4}$. In practice the unitarised limit has to be obtained
iteratively~\citep{ATL-PHYS-PUB-2014-007}.
\begin{figure}[htbp!]
    \centering
    \includegraphics[width=0.7\textwidth]{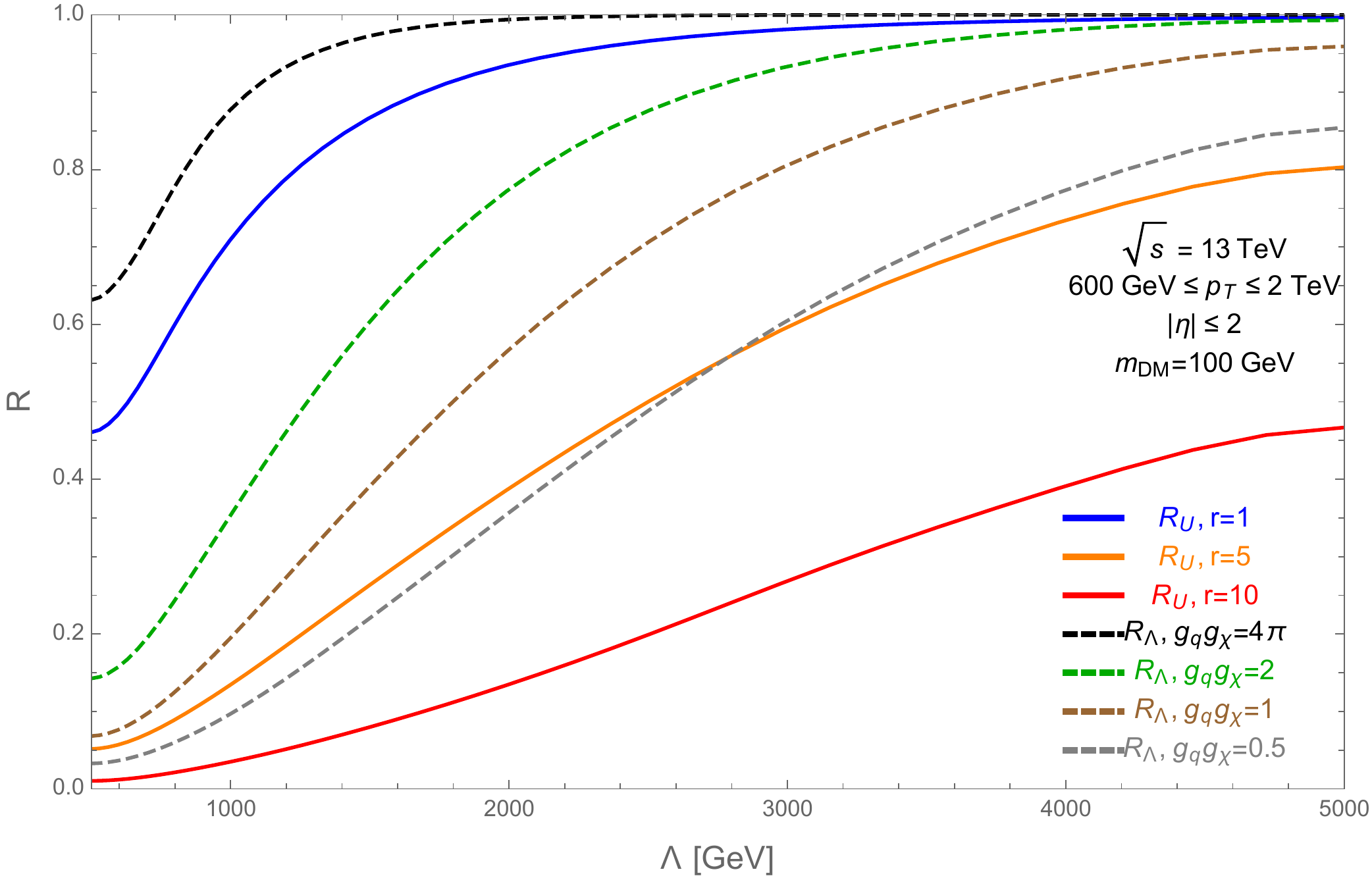}
    \caption{Quantities $R_{U},R_{\Lambda}$ as a function of the cut-off scale
    $\Lambda$ for different values of $r=1,5,10$ and $g_q g_\chi=0.5,1,2,4\pi$ for $m_{DM}=100 \GeV$. The solid lines refer to $R_{U}$, the dashed lines refer to $R_{\Lambda}$. Both gluon and quark jets were included in both cases.}
    \label{fig:A}
\end{figure}
\begin{figure}[htb!]
    \centering
    \includegraphics[width=0.7\textwidth]{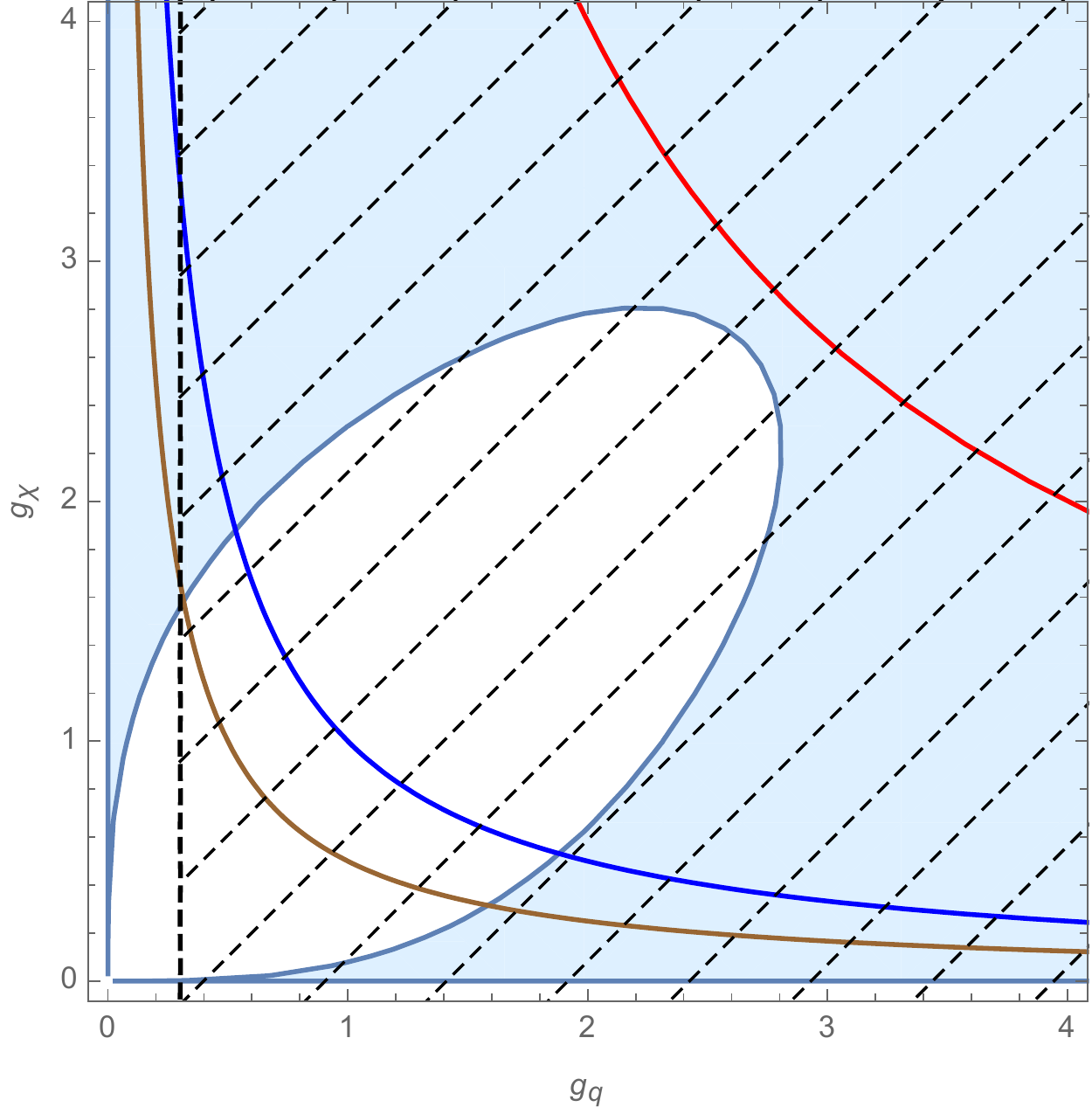}
    \caption{The light blue shaded region is the region of the
    parameter space with $R_U<R_\Lambda$. The region covered by black dashed lines is excluded by dijet search \citep{ATLAS:2015nsi} (for mediators below $3 \TeV$). The brown, blue and red lines are contours where the value of $g_q g_\chi$ is constant, and equal to $1/2$ (brown), $1$ (blue) and $8$ (red). 
    }
    \label{fig:region}
\end{figure}
Fig.~\ref{fig:A} shows the ratio $R_U$ as a function of the cutoff scale
$\Lambda$ for different values of $r=1,5,10$ as solid lines. The dashed lines
serve as a comparison to the corresponding ratio
\begin{align}
	R_\Lambda =
	\frac{\sigma_\mathrm{truncated,coll.}}{\sigma_\mathrm{EFT,coll.}}\;,
\end{align}
using the truncated amplitudes for different benchmark values of the couplings
$g_qg_\chi=0.5,1,2,4\pi$. All ratios have been obtained using the collinear
approximation including exactly one jet, which can be either a quark or a gluon jet. The centre of mass energy is fixed to 
$\sqrt{s}=13 \TeV$ and the DM mass to $m_{DM}=100 \GeV$. The transverse momentum
$p_T$ is limited to $600 \GeV \leq p_T \leq 2 \TeV$ and rapidity is required to
satisfy $|\eta|\leq 2$. The ratios $R_U$ and $R_\Lambda$ do not change much if
the cut on $p_T$ is slightly increased to $700 \GeV$. 

The suppression is generally stronger for low cut-off scales $\Lambda$, because more events
have to be discarded using the truncation procedure or the amplitude is reduced
for smaller center of mass energies $\sqrt{s}$ using $K$-matrix unitarisation.
The more a value deviates from $r=1$, the more the unitarised cross section is
suppressed, similar to smaller couplings $g_qg_\chi$ when using truncation. This
can be clearly seen in Fig.~\ref{fig:A}.

The values of $R_U$ reported in Fig.~\ref{fig:A} can be used to rescale EFT
limits in the same way as with $R_\Lambda$. The precise description of the
rescaling procedure and its main consequences are outlined in Ref.~\citep{ATL-PHYS-PUB-2014-007}.

Finally we compare the $K$-matrix unitarisation to the truncation procedure in the
Fig.~\ref{fig:region}. The solid lines show the lines of constant
$g_qg_\chi=0.5,1,8$ from left to right. The vertical dashed line indicates the
current limit from dijet searches restricting $g_q\lesssim 0.25$ for mediator
masses up to 3 TeV \citep{ATLAS:2015nsi}. 
The light blue shaded region has $R_U<R_\Lambda$, i.e.~unitarisation leads to
a larger suppression of the cross section than truncation and thus a less
stringent limit.
Generally the truncated amplitude is less suppressed for
$g_qg_\chi \gtrsim 3$ and thus leads to a stronger limit. In the region which is
not excluded by the dijet constraint, i.e. $g_q\lesssim 0.25$, we find 
that the unitarisation method leads to a stronger limit, $R_U>R_\Lambda$, for
$g_\chi\lesssim 1$. 

\section{Conclusions}
\label{sec:conclusions}
Non-renormalisable operators lead to violation of perturbative
unitarity in scattering amplitudes above the scale of the
operator. This particularly poses a problem for the interpretation of
monojet searches at the LHC experiments in terms of EFTs, because the
limits on the cut-off scale $\Lambda$ obtained assuming an EFT are
lower than the centre of mass energy $\sqrt{s}$. Thus there are many
high-energy collisions with a centre of mass energy greater than
$\Lambda$. Although high-energy events are penalised by the small
values of the parton distribution functions, this is cancelled by the
enhanced scattering amplitude, which grows proportional to
the centre of mass energy.

$K$-matrix unitarisation allows consistent limits to be obtained
within the EFT framework.  We exemplified this for the operator D5 as
well as two other simple toy models.  It leads to a smooth suppression
of the scattering amplitude. In the limit of large centre of mass
energy, $\sqrt{s}\to\infty$, the $T$-matrix approaches $i
\mathbb{1}$ and thus the off-diagonal elements describing DM pair production at
the LHC vanish. $K$-matrix unitarisation introduces a dependence on the other
$T$-matrix elements and thus the cut-off scales of other operators, e.g. four quark operators and
operators with four DM particles. The smallest cut-off scale among all relevant
operators determines the scale when the suppression due to $K$-matrix
unitarisation sets in. Hence the least suppression of the cross section in the
$K$-matrix unitarisation framework is obtained if the cut-off scales are of a
similar order of magnitude. This can be clearly seen for the D5 operator: The
suppression increase with $r=\Lambda_{q\chi}/\Lambda_{\chi\chi}$, since the smallest cut-off scale decreases with
$r$.

We recast the ATLAS 8 TeV monojet limit on the operator D5 for five
benchmark values of $r=1,2,3,4,5$ finding a slight suppression of a
few percent for $r=1$ which grew to more than $50 \%$ for $r=5$. Given
the suppression of the cross section as a function of the cut-off
scale $\Lambda$, it is straightforward to recast the limit obtained
using an EFT to a limit for the unitarised EFT. We provide this ratio
for three different choices of $r$, for a centre of mass energy of
$\sqrt{s}= 13 \TeV$, which can be directly used to obtain the 
unitarised EFT limit given the EFT limit.  Note however that all
results have been obtained in the collinear approximation and without
including a possible second jet. Going beyond these two approximations,
and the application of the same procedure to the other considered
operators, will be an interesting extension of the present work.

$K$-matrix unitarisation of EFT amplitudes provides a new way to
extract model-independent and theoretically reliable limits on the
dark matter production cross section at the LHC. The method can be
applied to a wide class of scenarios, including other mono-X searches
or Simplified Models without manifest gauge invariance, providing, in
certain cases, more stringent limits than the truncation method
currently used.

\vspace*{50px}

\textbf{Acknowledgements} 
We thank Lei Wu for collaboration during the initial stages of this project. 
This work was supported in part by the Australian Research Council.

%
\appendix
\section{Collinear Approximation}\label{app:collinear}
The collinear approximation allows to drastically simplify the discussion. This
appendix contains a detailed derivation of the relevant cross section.
Starting by simplifying the three-body phase space, we can write
\be
d\phi^{3body}
=(2\pi)^4\delta^4(\sum_{i=1}^3 p_i-p_0)\prod_{i=1}^3 \frac{d^3p_i}{(2\pi)^3
2E_i}
=\frac{1}{2^8\pi^5} \frac{d^3p_1 d^3p_2}{E_1 E_2 E_3}
\delta(E_1+E_2+E_3-E_0)\;.
\ee
The phase space is Lorentz invariant, so we are free to evaluate this expression
in any reference system. 
After introducing the four-momentum $p_{23}=p_2+p_3$ with the corresponding
energy $E_{23}=p_{23}^0$ and invariant four-momentum $s_{23}=p_{23}^2$, it is possible to use the identities
\begin{align}
	1 & = ds_{23}\delta(s_{23}-p_{23}^2)\theta(p_{23}^0)\label{eq:s23}\\
	\delta(E_1+E_2+E_3-E_0) & =\delta(E_2+E_3-E_{23})\delta(E_1+E_{23}-E_0)dE_{23}\label{eq:phdelta}
\end{align}
to separate the two-body phase space of particles $2$ and $3$
\begin{align}
d\phi^{3body}&=\frac{1}{2^8\pi^5}\frac{d^3p_1 d^3p_2}{E_1 E_2 E_3} ds_{23}
	\delta(s_{23}-p_{23}^2)\theta(p_{23}^0)\delta(E_2+E_3-E_{23})\delta(E_1+E_{23}-E_0)dE_{23}\\
&=\frac{1}{2^4\pi^3}\frac{d^3p_1}{E_1} ds_{23}
	\delta(s_{23}-p_{23}^2)\theta(p_{23}^0)\delta(E_1+E_{23}-E_0)dE_{23}d\phi^{2body}_{2,3}\;,
\end{align}
where in the last step we have used the definition of the two-body phase space
of the particles $2$ and $3$
\be
d\phi^{2body}_{2,3}\equiv\frac{1}{2^4\pi^2}\frac{d^3p_2}{E_2
E_3}\delta(E_2+E_3-E_{23})\;,
\ee
which will be included in the two-body cross section. The remaining part
can be further simplified by integrating over $E_{23}$ 
\begin{align}
d\phi^{3body}&=d\phi^{2body}_{2,3}\frac{ds_{23}}{2\pi}\frac{1}{2^4\pi^2}\frac{d^3p_1}{E_1
E_{23}}\delta(E_1+E_{23}-E_0)
\;.
\end{align}
While we are not interested in simplifying $d\phi^{2body}_{2,3}$ further, as
its expression in terms of kinematic variables will be necessary only to
calculate the cross section $\sigma_{q\bar{q}\rightarrow\chi\bar{\chi}}$, we
want to simplify the last delta function in 
\bea
d\phi^{3body}
&=&d\phi^{2body}_{2,3}\frac{ds_{23}}{2^4\pi^2}\frac{d\cos\theta_0 E_1
dE_1}{E_{23}}\delta(E_1+E_{23}-E_0)\;,
\eea
which can be evaluated using 
\begin{align}
	E_{23}&=\sqrt{E_1^2+s_{23}} &
	\frac{dE_{23}}{dE_1}&=\frac{E_1}{\sqrt{E_1^2+s_{23}}}=\frac{E_1}{E_{23}}\;.
\end{align}
Thus we obtain after the integration with respect to $E_1$
\be
d\phi^{3body}
=d\phi^{2body}_{2,3}\frac{ds_{23}d\cos\theta_0}{2^4\pi^2}\frac{E_1}{E_0}
\ee
The phase space and cross section are simple to evaluate in the centre of mass
frame, where momentum fraction of the partons equal $x_1=x_2=x$ and the following kinematic relations hold
\begin{align}
\hat{s}&=(p_1+p_2+p_3)^2=s x^2 &
E_1&=\sqrt{s x^2} \frac{z_0}{2} \\
	s_{23}&=(p_2+p_3)^2=s x^2 (1-z_0) &
E_{23}&=
	\sqrt{s x^2} \left(1-\frac{z_0}{2}\right)
\end{align}
The definition of $z_0,\theta_0$ is given in the following parametrisation of
the momenta in the centre of mass frame
\bea
p_1^\mu&=&\sqrt{s x^2} \frac{z_0}{2} \left(1,0,\sin\theta_0,\cos\theta_0\right)\\
p_2^\mu&=&\sqrt{s x^2} \left(\frac{1-y_0}{2},\sqrt{(1-y_0)^2-a^2}\hat{p}_3\right)\\
p_3^\mu&=&\sqrt{s x^2}
\left(\frac{1+y_0-z_0}{2},\sqrt{(1+y_0-z_0)^2-a^2}\hat{p}_4\right)\;,
\eea
where the angle between $p_2$ and $p_1$ is fixed by momentum conservation and 
the fraction $\tfrac{2m_{DM}}{\sqrt{s x^2}}$. 
Using this parametrisation allows us to write the three-body phase space as
\be
d\phi^{3body}=d\phi^{2body}_{2,3}\frac{s x^2 z_0}{32\pi^2} dz_0 d\cos\theta_0
\ee
clearly separating the two-body phase space factor from the variables $z_0$ and
$\cos\theta_0$ describing the additional jet.

After the derivation of the convenient form of the three-body phase space
factor, we are ready to work with the collinear approximation. The four-momentum of
the jet is denoted $p_1$, while the four-momenta of the DM particles are
$p_{2,3}$. 
Following the standard discussion of the collinear limit (See
e.g.~\cite{Peskin:1995ev}), the monojet cross section with a gluon jet can be
written as~\footnote{Note that only one of the two
	diagrams contributes, as only one can be "collinear". Consequently also
interference is negligible.}
\begin{align}
d\sigma_{q\bar{q}\rightarrow\chi\bar{\chi}+j(g)}
&=\frac{1}{|v_{q}-v_{\bar{q}}| 2E_q 2E_{\bar{q}}} 
\left[\frac12 \sum |M|^2\right] \frac{1}{(p_{q,\bar{q}}-p_1)^4} \frac{1}{4}
\overline{| M|^2}_{q\bar{q}\rightarrow\chi\bar{\chi}}(s_{23})
 d\phi^{3body}
\\\nonumber
&=\frac{1}{2sx^2} \left[\frac{2 g_s^2 p_T^2}{z_0(1-z_0)}
\frac{1+(1-z_0)^2}{z_0}\right] \frac{z_0^2}{p_T^4} \frac{1}{4}\overline{|M|^2}_{q\bar{q}\rightarrow\chi\bar{\chi}}(s_{23})
 d\phi^{2body}_{2,3}\frac{s x^2 z_0}{32\pi^2} dz_0 d\cos\theta_0
 \end{align}
 neglecting the color factor. The four-momentum $p_{q,\bar q}$ denotes the initial state four-momentum of the parton
radiating off the gluon and the transverse momentum of the gluon is given by
\begin{equation}\label{eq:pTjet}
	p_T = \sqrt{s x^2} \frac{z_0}{2} \sin\theta_0\;.
\end{equation}
The $2\to 2$ scattering cross section for $q\bar q\to\chi\bar\chi$, 
\begin{equation}
	\sigma_{q\bar q \to\chi\bar\chi} (s_{23}) = \frac14 \frac{\overline{\left| 
			M\right|^2}_{q\bar q\to\chi\bar \chi}(s_{23})}{2sx^2
		(1-z_0)} d \phi^{2body}_{2,3}\;,
	\end{equation}
can be factored out leading to 
  \begin{align}
  d\sigma_{q\bar{q}\rightarrow\chi\bar{\chi}+j(g)}
&=\sigma_{q\bar{q}\rightarrow\chi\bar{\chi}}(s_{23}) \frac{\alpha_s}{4\pi}
\frac{1+(1-z_0)^2}{z_0} \frac{z_0}{p_T^2} s x^2 z_0 dz_0
d\cos\theta_0\\
&=\sigma_{q\bar{q}\rightarrow\chi\bar{\chi}}(s_{23}) \frac{\alpha_s}{\pi}
\frac{1+(1-z_0)^2}{z_0} \frac{1}{\sin^2\theta_0}  dz_0 d\cos\theta_0\;,
\end{align}
where Eq.~\eqref{eq:pTjet} has been used in the last line.
Finally the cross section has to expressed in terms of the variables in the lab
frame to properly take the detector geometry into account. 
The change from the so-far considered variables in the centre of mass frame
$(z_0,\theta_0)$ to the transverse momentum and rapidity of the jet,
$(p_T,\eta)$, leads to the following Jacobian factor
\be
	\frac{ dz_0 d\cos\theta_0}{d p_T d\eta} = \frac {4 p_T} {s x_1 x_2 z_0} 
\ee
and the old variables can be rewritten as follows
\begin{align}
        \frac{1}{\sin^2\theta_0} & = \frac{s x_1 x_2}{4 p_T^2} z_0^2
					&
        z_0 &= \frac{p_T}{\sqrt{s}} \frac{x_1 e^{-\eta} +x_2 e^{\eta}}{x_1x_2}
	\;.
\end{align}
Finally the color factors have to be included. For gluon emission it is 
$1/3$ for color average, $\tr[T_a T_a]=1/2$ and a factor of $8$ for the sum over
gluons. Thus the color factor is $C_F=4/3$. The cross section $\sigma_{q\bar q
\to \chi\bar\chi}$ contains the color factor $1/3$: $(1/3)^2$ for the color average and $3$ for the color
sum. Thus the color factor for the full 3body cross section is $4/9$ and the final expression for the emission of a gluon jet replacing $x^2$ 
by $x_1x_2$ is given by
\be
\sigma_{q\bar{q}\rightarrow\chi\bar{\chi}+j(g)}=\sum_q\int dx_1 dx_2 dp_T d\eta 
(f_q(x_1)f_{\bar{q}}(x_2)+f_q(x_2)f_{\bar{q}}(x_1))
\frac{dzd\cos\theta_0}{dp_T d\eta}
\sigma_{q\bar{q}\rightarrow\chi\bar{\chi}}(s_{23}) P_{q\to g}(z_0,\theta_0)
\ee
with the splitting function
\be
P_{q\to g}(z_0,\theta_0)=\frac{4\alpha_s}{3\pi} \frac{1+(1-z_0)^2}{z_0
\sin^2\theta_0}\;.
\ee
This expression is consistent with the expression in Ref.~\citep{Birkedal:2004xn}.
The factor $2$ for the $2$ emissions from the initial quark and anti-quark lines is already taken into account,
because the expression is only valid for $\theta\in(0,\theta_{max})$ for the
emission from parton 1 or $\theta\in(\theta_{max},\pi)$ for the emission from
parton 2. Each time only one of the 2 diagrams contributes. Extending to the
maximum, i.e. $\theta_{max}=\pi/2$, the cross section is given by the calculated expression integrated over the full
range of $\theta$, without any additional factor of $2$.

Similarly, the cross section for radiating off a quark-jet is given by
\begin{multline}
\sigma_{q\bar{q}\rightarrow\chi\bar{\chi}+j(q)}=\sum_{q} \int dx_1 dx_2 dp_T d\eta 
\left(f_q(x_1) f_g(x_2) + f_q(x_2) f_g(x_1) + \left[q\to \bar q\right]
\right)\\ 
\frac{dzd\cos\theta_0}{dp_T d\eta}
\sigma_{q\bar{q}\rightarrow\chi\bar{\chi}}(s_{23}) P_{g\to q}(z_0,\theta_0)\;,
\end{multline}
where the splitting function for a quark-jet with $n_f$ different possible quark flavours is 
\be
P_{g\to q}(z_0,\theta_0)=\frac{n_f\, \alpha_s}{4\pi}
\frac{z_0^2+(1-z_0)^2}{\sin^2\theta_0}\;.
\ee

\section{Convention for Spinors} \label{eq:helicityspinors}
We explicitly list the helicity spinors used in our calculations to fix the
convention of phases. In the ultra-relativistic limit and setting the
azimuthal angle $\phi=0$, the helicity spinors take the form
\begin{align}
	u_R(E,\theta) & = v_L(E,\theta) = \sqrt{2 E} \begin{pmatrix} 0 \\ 0 \\ \cos\frac\theta2
\\ i \sin\frac\theta2 \end{pmatrix} & 
	u_L(E,\theta) & = - v_R(E,\theta) = \sqrt{2 E} \begin{pmatrix} i\sin\frac\theta2\\
	\cos\frac\theta2 \\ 0 \\ 0 \end{pmatrix}\;. 
\end{align}

\section{K-Matrix Unitarisation of D5}\label{app:D5} 
The $T$-matrix for $2\to 2$ scattering of quark - anti-quark and DM-DM two-particle states in case of the effective theory described
by the Lagrangian in Eq.~\eqref{eq:D5} is given by
\begin{equation}
	T=-\frac{1}{16\pi^2}
	\begin{pmatrix}
\frac{2 s \cos ^2\left(\frac{\theta }{2}\right)}{\Lambda _{qq}^2} & 0 & 0 & \frac{s
   \sin ^2\left(\frac{\theta }{2}\right)}{\Lambda _{qq}^2} & \frac{s \cos
   ^2\left(\frac{\theta }{2}\right)}{\Lambda _{q\chi}^2} & 0 & 0 & \frac{s \sin
   ^2\left(\frac{\theta }{2}\right)}{\Lambda _{q\chi}^2} \\
 0 & \frac{s}{\Lambda _{qq}^2} & 0 & 0 & 0 & 0 & 0 & 0 \\
 0 & 0 & \frac{s}{\Lambda _{qq}^2} & 0 & 0 & 0 & 0 & 0 \\
 \frac{s \sin ^2\left(\frac{\theta }{2}\right)}{\Lambda _{qq}^2} & 0 & 0 & \frac{2 s
   \cos ^2\left(\frac{\theta }{2}\right)}{\Lambda _{qq}^2} & \frac{s \sin
   ^2\left(\frac{\theta }{2}\right)}{\Lambda _{q\chi}^2} & 0 & 0 & \frac{s \cos
   ^2\left(\frac{\theta }{2}\right)}{\Lambda _{q\chi}^2} \\
 \frac{s \cos ^2\left(\frac{\theta }{2}\right)}{\Lambda _{q\chi}^2} & 0 & 0 &
   \frac{s \sin ^2\left(\frac{\theta }{2}\right)}{\Lambda _{q\chi}^2} & \frac{2 s
   \cos ^2\left(\frac{\theta }{2}\right)}{\Lambda _{\chi \chi }^2} & 0 & 0 & \frac{s \sin
   ^2\left(\frac{\theta }{2}\right)}{\Lambda _{\chi \chi }^2} \\
 0 & 0 & 0 & 0 & 0 & \frac{s}{\Lambda _{\chi \chi }^2} & 0 & 0 \\
 0 & 0 & 0 & 0 & 0 & 0 & \frac{s}{\Lambda _{\chi \chi }^2} & 0 \\
 \frac{s \sin ^2\left(\frac{\theta }{2}\right)}{\Lambda _{q\chi}^2} & 0 & 0 &
   \frac{s \cos ^2\left(\frac{\theta }{2}\right)}{\Lambda _{q\chi}^2} & \frac{s
   \sin ^2\left(\frac{\theta }{2}\right)}{\Lambda _{\chi \chi }^2} & 0 & 0 & \frac{2 s \cos
   ^2\left(\frac{\theta }{2}\right)}{\Lambda _{\chi \chi }^2} \\
	\end{pmatrix}
\end{equation}
in the basis 
$\left(
	\ket{q_L\bar q_R},
 	\ket{q_L\bar q_L},
	\ket{q_R\bar q_R},
	\ket{q_R\bar q_L},
	\ket{\chi_L\bar \chi_R},
	\ket{\chi_L\bar \chi_L},
	\ket{\chi_R\bar \chi_R},
	\ket{\chi_R\bar \chi_L}
\right)$. The two-particle states with the same helicity, completely decouple
from the other states and can be treated separately. They are pairwise related
by parity and they only contribute to the
$J=0$ term in the partial wave expansion 
\begin{align}
	\braket{q_L \bar q_L|T^0|q_L\bar q_L} & =
	\braket{q_R \bar q_R|T^0|q_R\bar q_R}=
	-\frac{1}{4\pi}\frac{s}{\Lambda_{qq}^2}\\
	\braket{\chi_L \bar \chi_L|T^0|\chi_L\bar \chi_L} & =
	\braket{\chi_R \bar \chi_R|T^0|\chi_R\bar \chi_R}=
	-\frac{1}{4\pi}\frac{s}{\Lambda_{\chi\chi}^2}\;.
\end{align}
Thus the only non-vanishing elements of the unitarised $T$-matrix, $T^0$, are
given by
\begin{align}
	\braket{q_L \bar q_L|T^0_U|q_L\bar q_L} & =
	\braket{q_R \bar q_R|T^0_U|q_R\bar q_R}=
	\frac{\iu s }{s-4\pi \iu \Lambda_{qq}^2}\\
	\braket{\chi_L \bar \chi_L|T^0_U|\chi_L\bar \chi_L} & =
	\braket{\chi_R \bar \chi_R|T^0_U|\chi_R\bar \chi_R}=
	\frac{\iu s }{s-4\pi \iu \Lambda_{\chi\chi}^2}\;.
\end{align}
The remaining states with opposite helicities 
contribute to the $J=1$ term in the partial wave expansion.
The $4\times 4$ sub-block of the $T$-matrix, $T^1$, in the basis
$\left(
	\ket{q_L\bar q_R},
	\ket{q_R\bar q_L},
	\ket{\chi_L\bar \chi_R},
	\ket{\chi_R\bar \chi_L}
\right)$ is given by
\begin{equation}
	T^1 = -\frac{1}{12\pi} \begin{pmatrix}
		\frac{2 s}{\Lambda_{qq}^2} & \frac{s}{\Lambda_{qq}^2} & \frac{s}{\Lambda_{q\chi}^2} & \frac{s}{\Lambda_{q\chi}^2} \\
 \frac{s}{\Lambda_{qq}^2} & \frac{2 s}{\Lambda_{qq}^2} & \frac{s}{\Lambda_{q\chi}^2} & \frac{s}{\Lambda_{q\chi}^2} \\
 \frac{s}{\Lambda_{q\chi}^2} & \frac{s}{\Lambda_{q\chi}^2} & \frac{2 s}{\Lambda_{\chi\chi} ^2} & \frac{s}{\Lambda_{\chi\chi} ^2} \\
 \frac{s}{\Lambda_{q\chi}^2} & \frac{s}{\Lambda_{q\chi}^2} & \frac{s}{\Lambda_{\chi\chi} ^2} & \frac{2 s}{\Lambda_{\chi\chi} ^2} \\
	\end{pmatrix}\;.
\end{equation}
Many of the elements are related by the time-reversal symmetry and
parity.~\cite{Goldberger:1960md,Martin:1970ept} There are only $6$ independent
matrix elements and we find for the independent elements of the unitarised $T$-matrix $T^1_U$
\begin{align}
	\braket{q_L\bar q_R|T^1_U|q_L\bar q_R}&=\frac{s \left(48 \pi
	\Lambda_{q\chi}^2 r^4 s+\iu r^2 \left(5 s^2-288 \pi ^2
	\Lambda_{q\chi}^4\right)+36 \pi  \Lambda_{q\chi}^2 s\right)}{\left(s-12
	\iu \pi  \Lambda_{q\chi}^2 r^2\right) \left(-36 \iu \pi
	\Lambda_{q\chi}^2 r^4 s+r^2 \left(5 s^2-144 \pi ^2
	\Lambda_{q\chi}^4\right)-36 \iu \pi  \Lambda_{q\chi}^2 s\right)}
	\\
	\braket{q_L\bar q_R|T^1_U|q_R\bar q_L}&=\frac{12 \pi  \Lambda_{q\chi}^2
	r^2 s \left(r^2 s-12 \iu \pi  \Lambda_{q\chi}^2\right)}{\left(s-12 \iu \pi
		\Lambda_{q\chi}^2 r^2\right) \left(-36 \iu \pi
\Lambda_{q\chi}^2 r^4 s+r^2 \left(5 s^2-144 \pi ^2 \Lambda_{q\chi}^4\right)-36
\iu \pi  \Lambda_{q\chi}^2 s\right)}
	\\
	\braket{q_L\bar q_R|T^1_U|\chi_L\bar \chi_R}&=
-\frac{12 \pi  \Lambda_{q\chi}^2 r^2 s}{36 \iu \pi  \Lambda_{q\chi}^2 r^4 s+r^2
\left(144 \pi ^2 \Lambda_{q\chi}^4-5 s^2\right)+36 \iu \pi  \Lambda_{q\chi}^2 s}
	\\
	\braket{q_L\bar q_R|T^1_U|\chi_R\bar \chi_L}&=
	\braket{q_L\bar q_R|T^1_U|\chi_L\bar \chi_R}\\
	\braket{\chi_L\bar \chi_R|T^1_U|\chi_L\bar \chi_R}&=\braket{q_L\bar
q_R|T^1_U|q_L\bar q_R} \left[r\to \frac1r \right]
	\\
	\braket{\chi_L\bar \chi_R|T^1_U|\chi_R\bar \chi_L}&=\braket{q_L\bar
q_R|T^1_U|q_R\bar q_L} \left[r\to \frac1r \right]
	\;.
\end{align}
The fourth equation follows from the interaction being vector-like and the last two
equations follow from the symmetry $q\leftrightarrow \chi$.
The remaining matrix elements can be obtained from time-reversal and parity
symmetry:
Time reversal symmetry implies that $T_U^1$ is symmetric, i.e. $T^1_U =
\left(T^1_U\right)^T$. Parity conservation implies that matrix elements are
invariant under flipping all helicities, i.e. 
$\braket{\lambda_1^\prime\lambda_2^\prime | T^1_U |
\lambda_1\lambda_2}=\braket{-\lambda_1^\prime-\lambda_2^\prime | T^1_U |
-\lambda_1-\lambda_2}$.

\section{Effective SM-WIMP Operators}
We list the operators coupling the SM to Dirac fermion WIMPs~\cite{Goodman:2010ku} in Tab.~\ref{tab:opList}.
\begin{table}[h]
\begin{tabular}{lccccccc}
\toprule
   Name    &
D1 &
D2 &
D3 &
D4 &
D5
\\
\midrule
Op. &
$\bar{\chi}\chi\bar{q} q$ & 
 $\bar{\chi}\gamma^5\chi\bar{q} q$ & 
 $\bar{\chi}\chi\bar{q}\gamma^5 q$ &
 $\bar{\chi}\gamma^5\chi\bar{q}\gamma^5 q$ & 
 $\bar{\chi}\gamma^{\mu}\chi\bar{q}\gamma_{\mu} q$  \\
 \midrule\midrule
Name  &
D6 &
D7 &
D8 &
D9 &
D10 
 \\
\midrule
Op&
 $\bar{\chi}\gamma^{\mu}\gamma^5\chi\bar{q}\gamma_{\mu} q$ & 
 $\bar{\chi}\gamma^{\mu}\chi\bar{q}\gamma_{\mu}\gamma^5 q$ &
$\bar{\chi}\gamma^{\mu}\gamma^5\chi\bar{q}\gamma_{\mu}\gamma^5 q$ & 
 $\bar{\chi}\sigma^{\mu\nu}\chi\bar{q}\sigma_{\mu\nu} q$ & 
 $\bar{\chi}\sigma_{\mu\nu}\gamma^5\chi\bar{q}\sigma_{\alpha\beta}q$ &
\\
\midrule\midrule
Name  &
D11 &
D12 &
D13 &
D14 & \\
\midrule
Op&
 $\bar{\chi}\chi G_{\mu\nu}G^{\mu\nu}$ &
 $\bar{\chi}\gamma^5\chi G_{\mu\nu}G^{\mu\nu}$ &
 $\bar{\chi}\chi G_{\mu\nu}\tilde{G}^{\mu\nu}$ &
 $\bar{\chi}\gamma^5\chi G_{\mu\nu}\tilde{G}^{\mu\nu}$ &
 \\
\bottomrule
\end{tabular}
\caption{Operators coupling SM to WIMPs first shown in Ref.~\cite{Goodman:2010ku}.}
\label{tab:opList}
\end{table}

\label{Bibliography}

\lhead{\emph{Bibliography}} 

\bibliography{Bibliography} 
 
\end{document}